\documentclass[english]{emulateapj}
\usepackage{epsfig}
\usepackage{amsmath}
\usepackage{array,multirow}
\usepackage{bm}
\numberwithin{equation}{subsection}
\usepackage{natbib}
\usepackage{amssymb}
\usepackage{amsmath}
\providecommand{\e}[1]{\ensuremath{\times 10^{#1}}}
\usepackage{longtable}
 \LongTables
\usepackage[flushleft]{threeparttable}

\def \rsub {$r_{\text{sub}}$}

\usepackage[T1]{fontenc}
\usepackage{babel}

\usepackage{array}
\usepackage{booktabs}
\usepackage{multirow}
\newcommand{\head}[1]{\textnormal{\textbf{#1}}}

\begin{document}

\title{A Model for Type 2 Coronal Line Forest (CL\lowercase{i}F) AGN}

\author{Ana Glidden\altaffilmark{1}, Marvin Rose\altaffilmark{2}$^{,}$\altaffilmark{3}, Martin Elvis\altaffilmark{2}, and Jonathan McDowell\altaffilmark{2}} 
\altaffiltext{1}{Massachusetts Institute of Technology, 77 Massachusetts Ave, Cambridge, MA 02139, USA; \texttt{aglidden@mit.edu}}
\altaffiltext{2}{Harvard-Smithsonian Center for Astrophysics, 60 Garden Street, Cambridge, MA 02138, USA; \texttt{melvis@cfa.harvard.edu}}
\altaffiltext{3}{Department of Physics and Astronomy, University of Sheffield, Sheffield S3 7RH, UK; \texttt{m.rose@sheffield.ac.uk}}

%%%%%%%%%%%%%%%%%%%%%%%%%%%% ABSTRACT %%%%%%%%%%%%%%%%%%%%%%%%%%%%
\begin{abstract}

We present a model for the classification of Coronal-Line Forest Active Galactic Nuclei (CLiF AGN). CLiF AGN are of special interest due to their remarkably large number of emission lines, especially forbidden high ionization lines (FHILs). \cite{Rose} suggest that their emission is dominated by reflection from the inner wall of the obscuring region rather than direct emission from the accretion disk. This makes CLiF AGN laboratories to test AGN-torus models. Modeling AGN as an accreting supermassive black hole, surrounded by a cylinder of dust and gas, we show a relationship between viewing angle and the revealed area of the inner wall. From the revealed area, we can determine the amount of FHIL emission at various angles. We calculate the strength of [\ion{Fe}{7}]$\lambda$6087 emission for a number of intermediate angles (30$^{\circ}$, 40$^{\circ}$, and 50$^{\circ}$) and compare the results with the luminosity of the observed emission line from six known CLiF AGN. We find that there is good agreement between our model and the observational results. The model also enables us to determine the relationship between the type 2 : type 1 AGN fraction vs the ratio of torus height to radius, $h/r$.

\end{abstract}

%%%%%%%%%%%%%%%%%%%%%%%%%%%% INTRODUCTION %%%%%%%%%%%%%%%%%%%%%%%%%%%%
\section{INTRODUCTION}

Supermassive black holes (SMBHs) are thought to lie at the centers of all galaxies. In some galaxies, the SMBH accretes matter rapidly, releasing enough gravitational potential energy as light to outshine its host galaxy. Yet, emission from this region, known as the Active Galactic Nucleus (AGN), is obscured to varying degrees for most active galaxies \citep{Risaliti,Lawrence2}. 

This obscuring region has long been modeled as a dusty torus that is both physically and optically thick \citep{Lawrence1, AntonucciMiller}. The angle of the torus relative to the line of sight (LOS) may then account for differences in observed AGN spectra. According to this Unification Model, emission from type 1 AGN comes directly from the accretion disk, while emission from type 2 AGN is viewed through the obscuring torus antiscattered into our LOS, producing polarized features \citep{Antonucci}. 

Developing the work of \citet{Nagao2000}, \defcitealias{Rose}{RET15}\citet[][hereafter RET15]{Rose} presented a new class of AGN called Coronal Line Forest (CLiF) AGN. Unlike the vast majority of AGN, CLiF AGN have dozens of forbidden high ionization lines (FHILs) with ionization potential > 54.4eV and large equivalent widths (EW > 5{\AA}), such as [\ion{Fe}{5}]$\lambda$3839, [\ion{Fe}{6}]$\lambda\lambda$5335,5426, [\ion{Fe}{7}]$\lambda$6087, [\ion{Fe}{10}]$\lambda$6375, [\ion{Ne}{5}]$\lambda$3426, and [\ion{Ar}{5}]$\lambda$7006 (\citetalias{Rose}). CLiF AGN are divided into two subcategories: type 2 CLiF AGN and type 1 AGN with strong coronal emission lines. 100 CLiF AGN, mainly type 1 with a few type 2, are now known~ (Rose et al. in prep). The broad and blue shifted kinematics of the FHILs in the type 1 CLiFs suggest a wind origin \citepalias{Rose}.

Here we seek to provide a model only for the subclass of type 2 CLiF AGN. \citetalias{Rose} suggested that the unique spectral properties of type 2 CLiF AGN likely originate from ablation of the inner wall of the torus by the central radiation source. To support this, \citetalias{Rose} showed that the coronal line emitting region lies at a distance approximately equal to the dust sublimation radius, the boundary of the inner torus wall \citep{Suganuma}. \citet{Rose2015b} employed the \citet{Fischer} relation between inclination angle and mid-IR color to show that type 2 CLiF AGN lie at angles intermediate between type 1 and type 2 AGN. 

Here we model this scenario for type 2 CLiF AGN in more detail to evaluate whether the inner torus wall could be the origin of FHIL emission.

%%%%%%%%%%%%%%%%%%%%%%%%%%%% MODELING %%%%%%%%%%%%%%%%%%%%%%%%%%%%
\section{Geometrical Constraints}

\begin{figure*}
  \begin{center}
    \leavevmode
	\includegraphics[width=8.7cm]{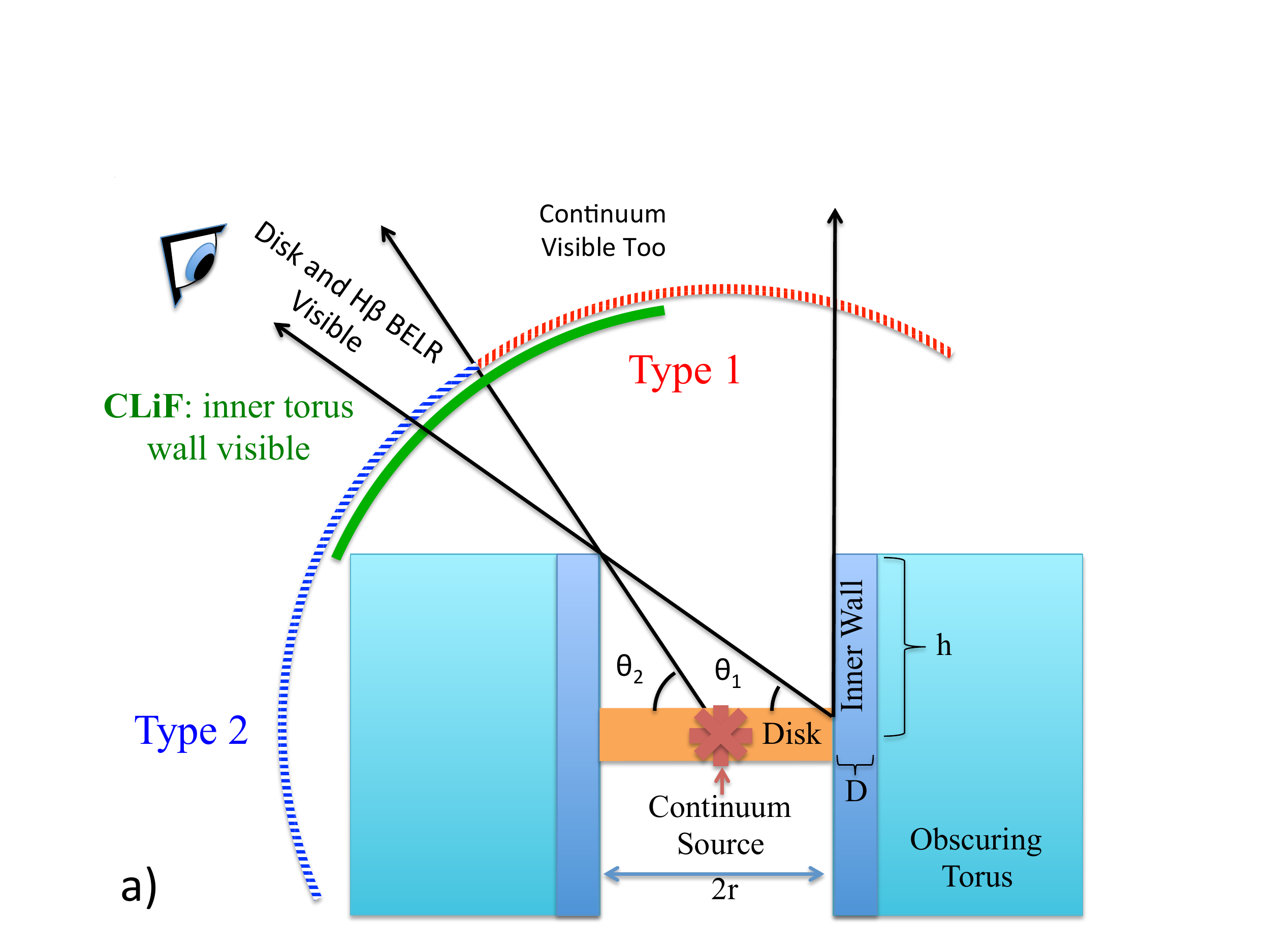} 
        \includegraphics[width=4.3cm]{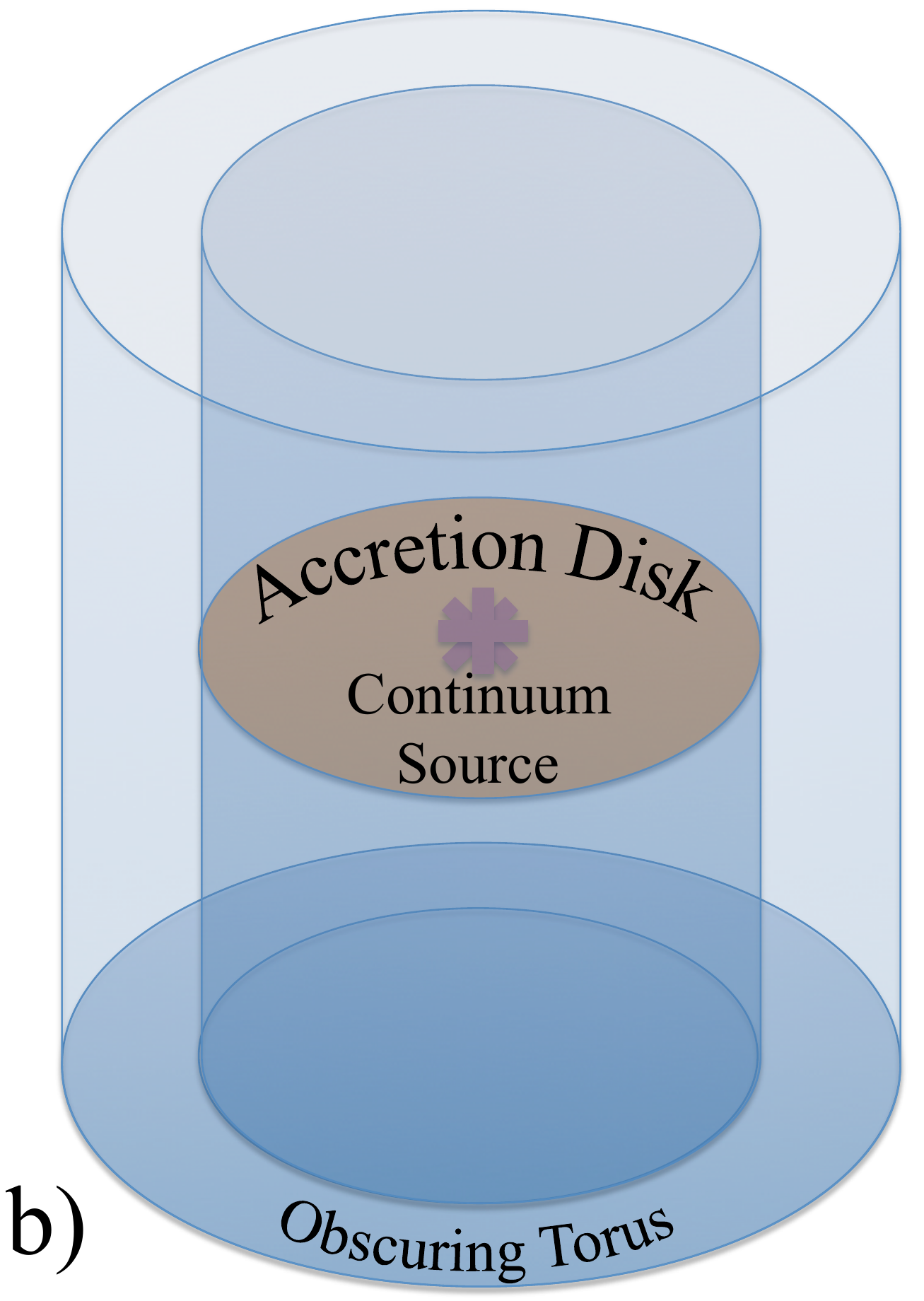}\\ 
       \caption[Cylinder]{(a) Cross Section of the Cylinder Model.  (b) 3D Diagram of the Cylinder Model. The height of the obscuring torus is labeled as $h$, the radius of the disk is $r$, and the thickness of the CLiF emitting region, $D$, is shown, but is not to scale.}
     \label{fig:cylinder1}
  \end{center}
\end{figure*}

\begin{figure*}
  \begin{center}
    \leavevmode
      \includegraphics[width=13cm]{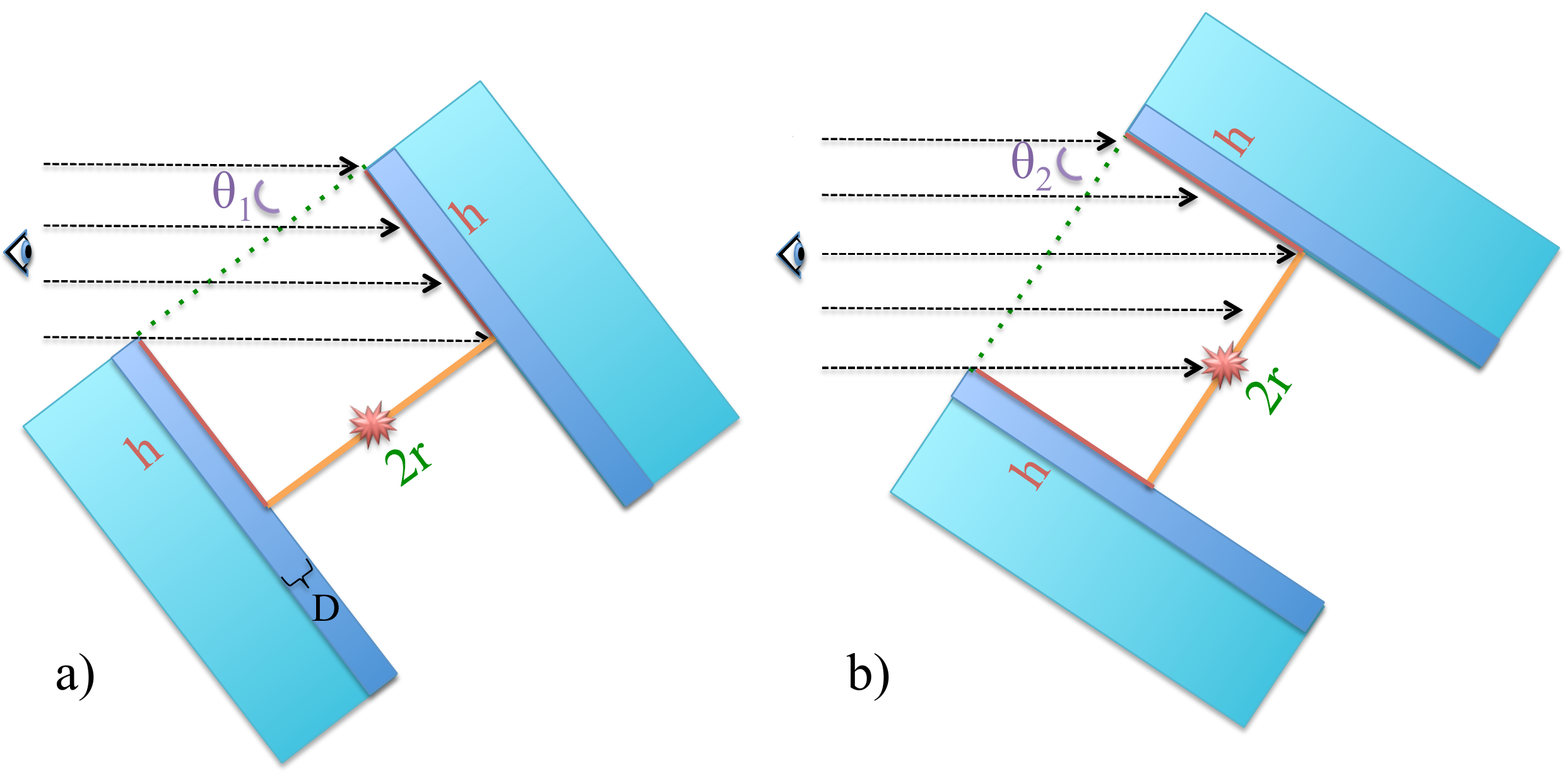}\\
       \caption[Cylinder]{Slice of the Cylinder Model showing what portion of the inner wall and the accretion disk can be seen at the two critical angles.  (a) The critical angle $\theta_{1}=\arctan(\frac{h}{2r})$, past which the accretion disk is revealed.  (b) The critical angle $\theta_{2}=\arctan(\frac{h}{r})$, past which the continuum source is revealed, outshining the emission from the inner wall. The thickness of the CLiF emitting region, $D$, is also shown, but is not to scale.}
     \label{fig:cylinder2}
  \end{center}
\end{figure*}

Our "cylinder model" comprises an optically and physically thick cylinder of gas and dust surrounding a physically thin, optically thick, central accretion disk \citep{Shields,Shakura,Malkan1982,Malkan1983} as shown in Figure \ref{fig:cylinder1}. The opaque disk extends across the entire width of the cylinder and is concentric with it. This disk blocks out emission from the lower part of the duty torus. A 50\% or greater opaque mid-plane disk is needed for three of the five AGN BLRs modeled by \citet{Pancoast}. This geometric assumption allows for flexibility between the various structures for the region between the torus and the SMBH that have been proposed. Whether this opaque mid-plane is due to the accretion disk, the BLR, or other structures \citep{Koshida,Fausnaugh,Lira}, their effect is the same within our model. For the purpose of the proposed model, its key role is in how it affects the area of the exposed inner wall. We also show the results without the disk. The radii of both the inner wall of the cylinder and the outer edge of the accretion disk are of length $r$. The height of the cylinder is $2h$ such that the height from the top of the obscuring region to the outer edge of the accretion disk is $h$ (Figure \ref{fig:cylinder2}.) The lower half of the torus does not contribute to the observed spectrum as it is blocked by the optically thick accretion disk. At the center is a point source of continuous emission.

%%%%%%%%%%%%%%%%%%%%%%%%%%%% Type2:Type1 %%%%%%%%%%%%%%%%%%%%%%%%%%%%
\subsection{Type 2 : Type 1 AGN Ratio and $h/r$}

\begin{figure}
  \begin{center}
    \leavevmode
     \includegraphics[width=7.2cm]{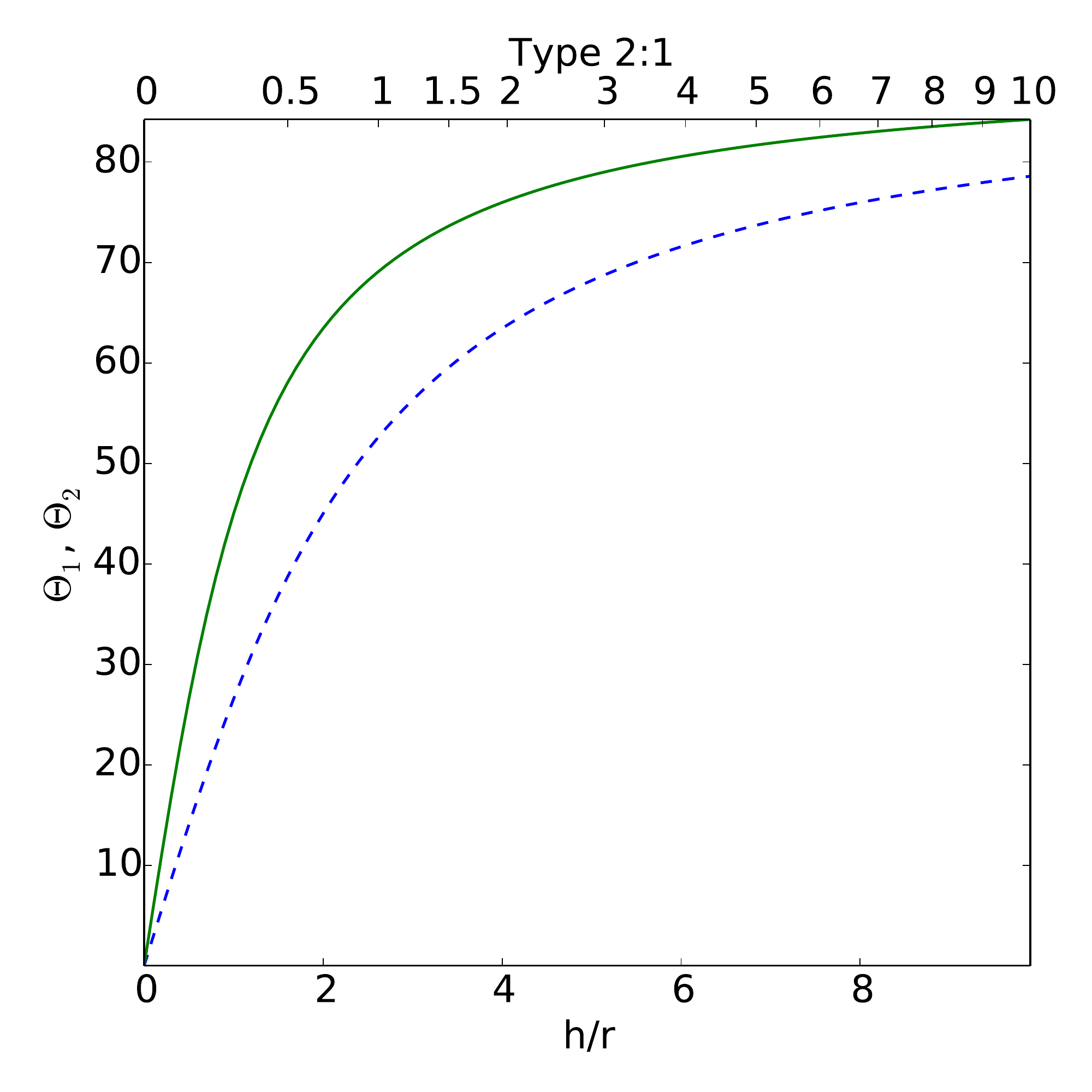}\\ 
     \includegraphics[width=7.2cm]{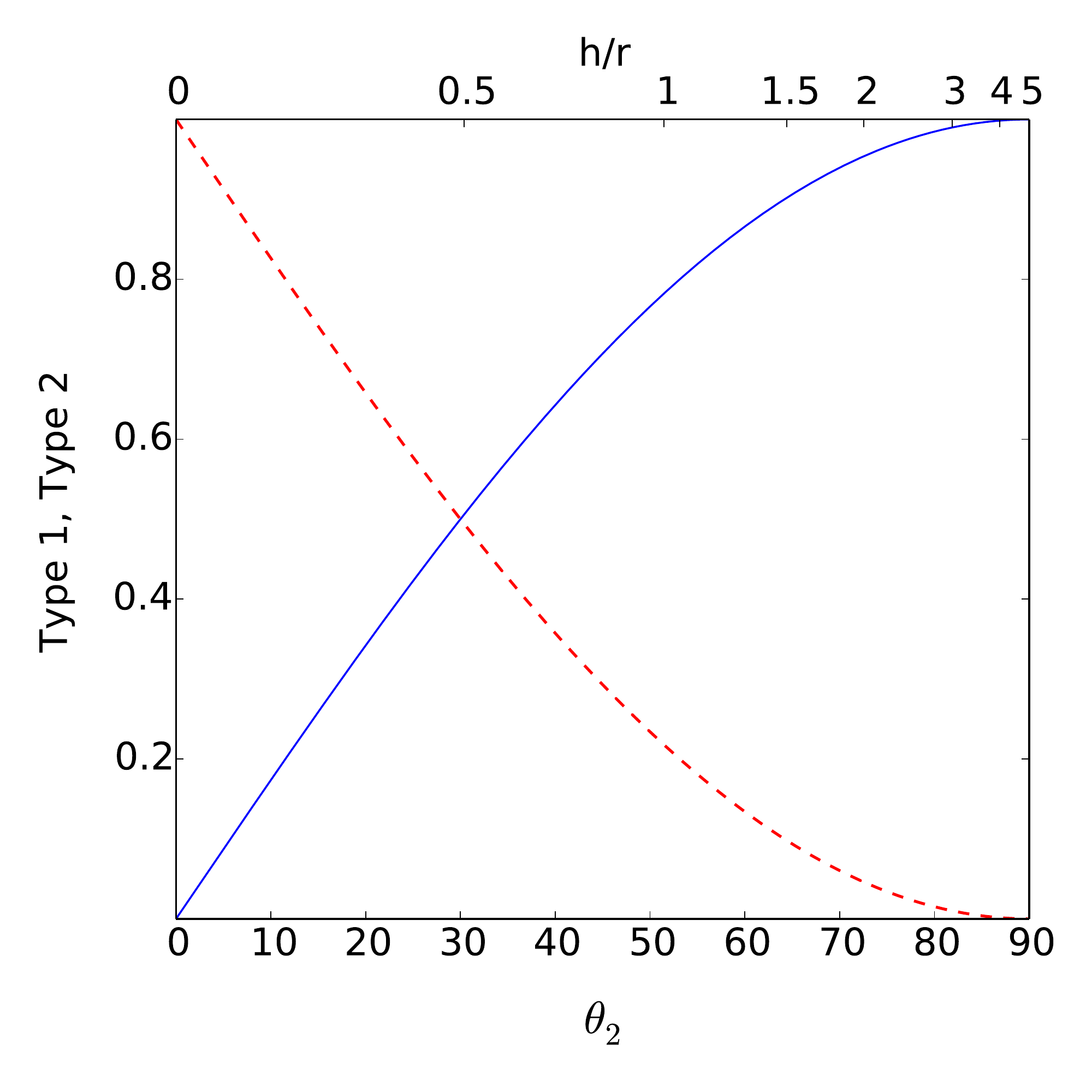} \\ 
     \includegraphics[width=7.2cm]{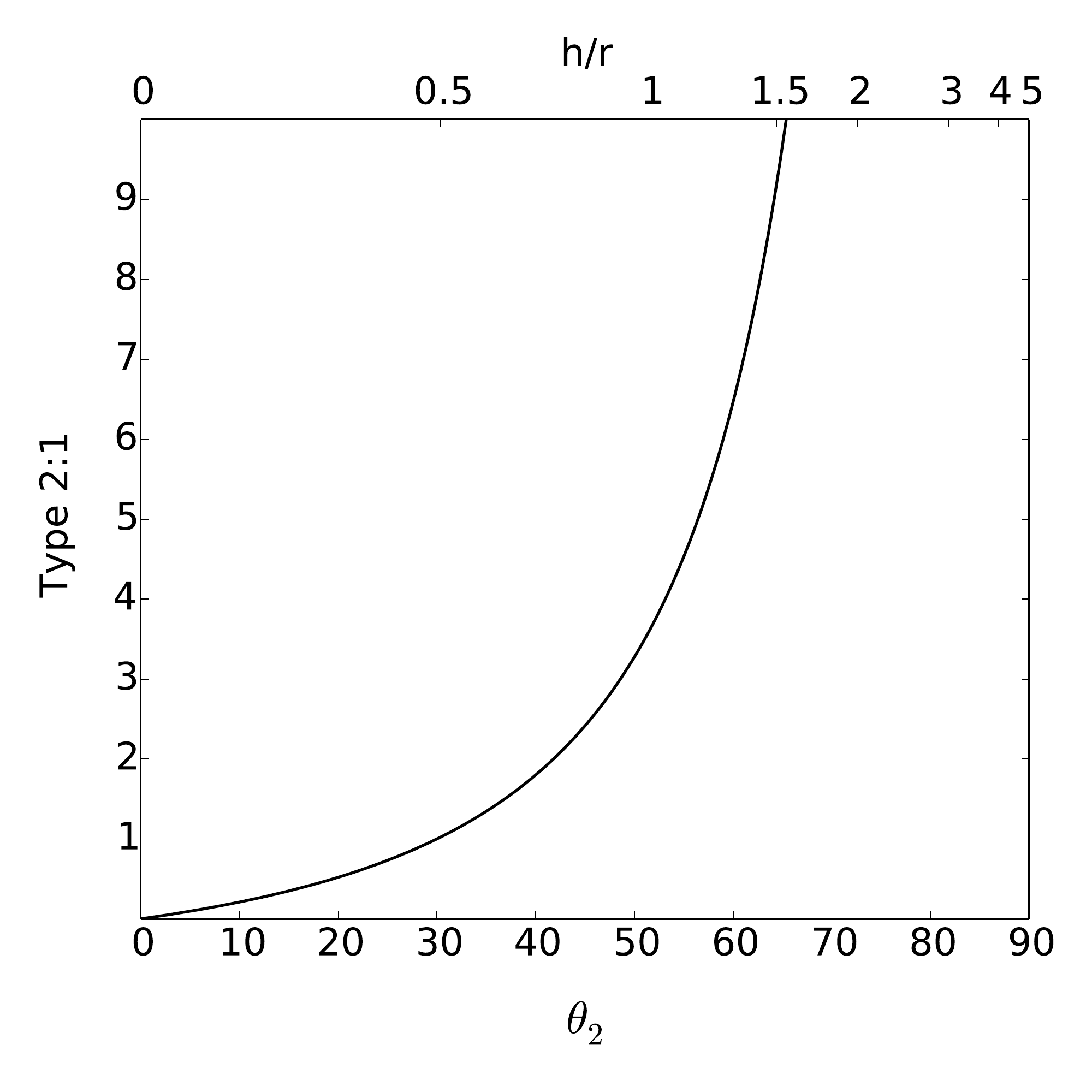}\\
       \caption[Ratio]{\textit{Top:} Dependence of $\theta_1$ (blue-dashed, where disk blocking begins) and  $\theta_2$ (green-solid, where the central engine becomes visible) on values of $h/r$ and the fraction of type 2 : type 1 AGN. \textit{Center:} The fraction of type 1 (red-dashed) and type 2 (blue-solid) AGN versus $\theta_{2}$ and ratio $h/r$. \textit{Bottom:} The fraction of type 2 : type 1 (black-solid) AGN as a function of the $h/r$ ratio. Table \ref{table:hrratios} lists these values for some cases of interest.}
     \label{fig:tri}
  \end{center}
\end{figure}

A given value of $h/r$ sets the critical viewing angle, $\theta_{2}$, where the central engine becomes visible and determines the type 2 : type 1 AGN ratio (Figure \ref{fig:cylinder2}b). In order to relate $h/r$ to the type 2 : type 1 AGN ratio, we first develop a formula for $\theta_{2}$.

As shown in Figure \ref{fig:cylinder1}b, looking edge-on ($\theta=0^{\circ}$) through the cylinder, neither the inner wall nor the accretion disk can be seen.  Observed face-on ($\theta=90^{\circ}$), only the accretion disk is seen.  In between edge-on and face-on, there exists a range of inclinations: $$\theta_{0}=0<\theta<\theta_{1}=\arctan(\frac{h}{2r})$$ where the inner wall of the cylinder can be seen while the accretion disk and BLR are still obscured. 

Past the critical angle of $\theta_{1}$, the inner wall and the accretion disk are both seen until a second critical angle: $$\theta_{2}=\arctan(\frac{h}{r})$$ is obtained, where the central region becomes visible, diluting the emission from the inner wall. Using $\theta_{2}$ we can calculate how the type 2 : type 1 AGN ratio changes as a function of $h/r$.

The normalized solid angle for type 1 AGN can be calculated to give the fraction of type 1 AGN: 

$$f_{1}=\frac{\int_{\theta_{2}}^{\frac{\pi}{2}} \! 2 \pi \rho[\theta] \, \mathrm{d}\rho}{\int_{0}^{\frac{\pi}{2}} \! 2 \pi \rho[\theta] \, \mathrm{d}\rho}=1-\sin\theta_{2}=1-\frac{1}{\sqrt{1+(r/h)^{2}}},$$
 where $\rho[\theta]=R\cos\theta$, $R=\sqrt{h^{2}+r^{2}}$, and $\theta_{2}=\frac{\pi}{2}-\arctan(\frac{r}{h}).$  Then, the fraction of type 2 AGN can be calculated from one minus the fraction of type 1 AGN: $$f_{2}=\frac{1}{\sqrt{1+(r/h)^2}}=\sin\theta_{2}.$$ Thus, the ratio of type 2 : type 1 AGN: $$\frac{f_{2}}{f{1}}=\frac{1}{\sqrt{1+(r/h)^2}-1}=\frac{1}{\csc\theta_{2}-1}.$$ These values are shown in Table \ref{table:hrratios} and plots showing these relations can be found in Figure \ref{fig:tri}.

Table \ref{table:hrratios} and  Figure \ref{fig:tri} show that the type 2 : type 1 AGN ratio is highly sensitive to the value of $\theta_{2}$ and so to $h/r$. In Figure \ref{fig:tri}a, the values of $\theta_{1}$ (blue-dashed) and $\theta_{2}$ (green-solid) can be seen as a function of $h/r$ and the type 2 : type 1 AGN ratio. Figure \ref{fig:tri}b shows how type 2 (blue-solid) and type 1 (red-dashed) AGN vary with $\theta_{2}$ and $h/r$. The relation between the ratio of the type 2 : type 1 AGN to $\theta_{2}$ and $h/r$ can be seen in Figure \ref{fig:tri}c.

Observations indicate type 2 : type 1 AGN ratios between 1:1 \citep{Lawrence2} using radio, mid/far-IR, optical, and hard X-ray and 4:1 \citep{Gilli} using soft and hard X-rays. Ratios between 1 and 4 imply a quite narrow range of $h/r$ from $\sim$0.6 to $\sim$1.3 and a correspondingly narrow range of $\theta_{2} \approx 30^{\circ} - 50^{\circ}$ (Figure \ref{fig:tri}, Table \ref{table:hrratios}). Hence, the discrepancies between different estimates of type 2 : type 1 AGN ratios may be due to quite small physical changes of $h/r$ that may arise from sample selection methods. 

%%%%%%%%%%%%%%%%%%%%%%%%%%%%%%%%%%%%%%%%%%%%%%%%%%%%%%%%%%%%%%%%%%%%%%%%%%%%%%%%%

\subsection{Torus Inner Radius}

In order to later calculate a FHIL luminosity, we need the ionizing flux at the dust sublimation radius, \rsub, where the inner boundary of the torus lies. \citetalias{Rose} observed values of $0.1<$\rsub$<4.3$ pc. We have chosen to use $h/r$ = 0.5, 1.0, and 1.5 in this paper. Choosing $r=1$ pc, we use $h=$ 0.5, 1.0, and 1.5 pc. 

Using the sublimation radius: $$ \text{\rsub}=0.4(\frac{L}{10^{45}\text{ erg s}^{-1}})^{1/2}(\frac{1500\text{ K}}{T_{sub}})^{2.6} \text{ pc}$$ from \citet{Elitzur} and $T_{sub}=1500$ K, the implied luminosity for \rsub=1 pc is 6\e{45} erg s$^{-1}$, similar to an AGN at the Seyfert/quasar boundary. An Eddington luminosity of this strength corresponds to a black hole mass of 5\e{7} M$_{\sun}$ \citep{Shankar}.

\begin{table}
\centering
\caption{Type 2 : Type 1 Ratios}
\begin{tabular}{ccc}
\toprule[0.75pt]
\toprule[0.25pt]
\multicolumn{1}{l}{\head{Type 2 : 1}} & \multicolumn{1}{c}{\head{$\theta_{2}$}}  & \multicolumn{1}{c}{\head{$h/r$}} \\
\hline
%\toprule[0.25pt]
1 : 1 & 30 & 0.58 \\
2 : 1 & 42 & 0.89 \\
3 : 1 & 49 & 1.13 \\
4 : 1	& 53 & 1.33 \\
10 : 1  & 65 & 2.18 \\
20 : 1 & 72 & 3.12 \\
65 : 1 & 80 & 5.68 \\
\hline
\end{tabular}
\label{table:hrratios}
\end{table}

%%%%%%%%%%%%%%%%%%%%%%%%%%%%%%%%%%%%%%%%%%%%%%%%%%%%%%%%%%%%%%%%%%%%%%%%%%%%%%%%%

\subsection{CLiF Inclination Angles}
Using high spatial resolution spectroscopy, \citet{Fischer13} estimated AGN inclinations for nearby (z < 0.06) type 1 and 2 AGN from the biconical outflow kinematics. The computed inclinations of the type 1 AGN studied in \citet{Fischer13} are clearly pole on (<i>=15+/-5 degrees) when compared to type 2 AGN (<i>=63+/-4 degrees). Interestingly, there is a lack of AGN with inclinations that are intermediate between these distributions \citep{Rose2015b}. Using WISE (W2-W4) colors as a proxy for the AGN inclination, Rose et al. (2015b) showed that type 2 CLiF AGN seem to have inclinations which are intermediate between typical type 1 and 2 AGN. Given this observation, we use $30^{\circ} - 50^{\circ}$ as CLiF inclinations in our model. This range of angles is similar to those found for the transitional range between the type 1 and 2 AGN studied in \cite{Marin}.

%%%%%%%%%%%%%%%%%%%%%%%%%%%%%%%%%%%%%%%%%%%%%%%%%%%%%%%%%%%%%%%%%%%%%%%%%%%%%%%%%%%%%%%%%%%%

\subsection{Inner Wall Area}
Next, we use the cylinder model to calculate the visible area of the inner wall as a function of viewing angle along our LOS: $A_{LOS}$. This area will allow us to calculate the strength of emission of the FHILs emanating from the inner wall.

\subsubsection{Optically Thick Disk Hidden: $\theta_{0}<\theta<\theta_{1}$}
\begin{figure*}
  \begin{center}
    \leavevmode
      \includegraphics[width=3.5cm]{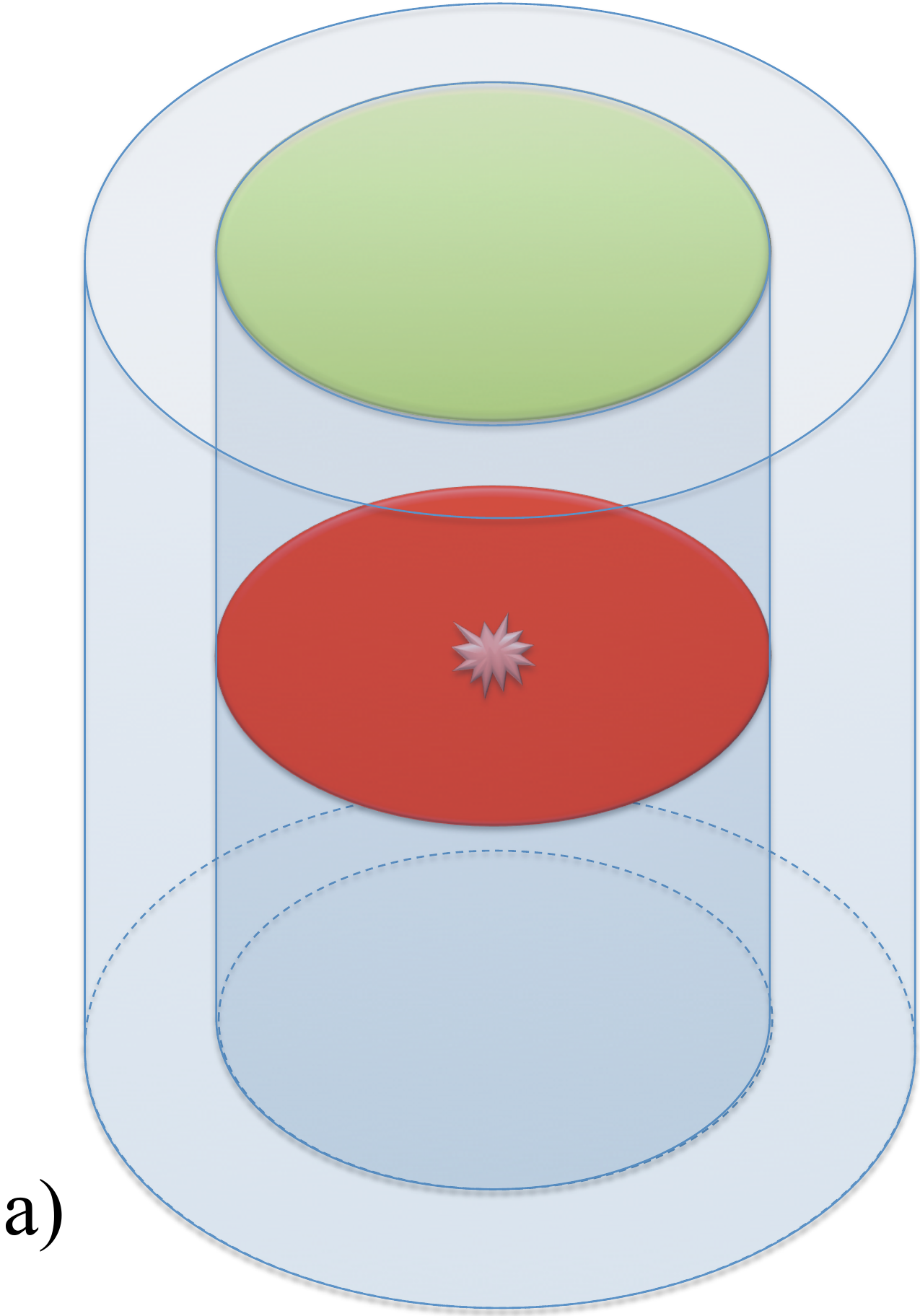} 
\includegraphics[width=3.5cm]{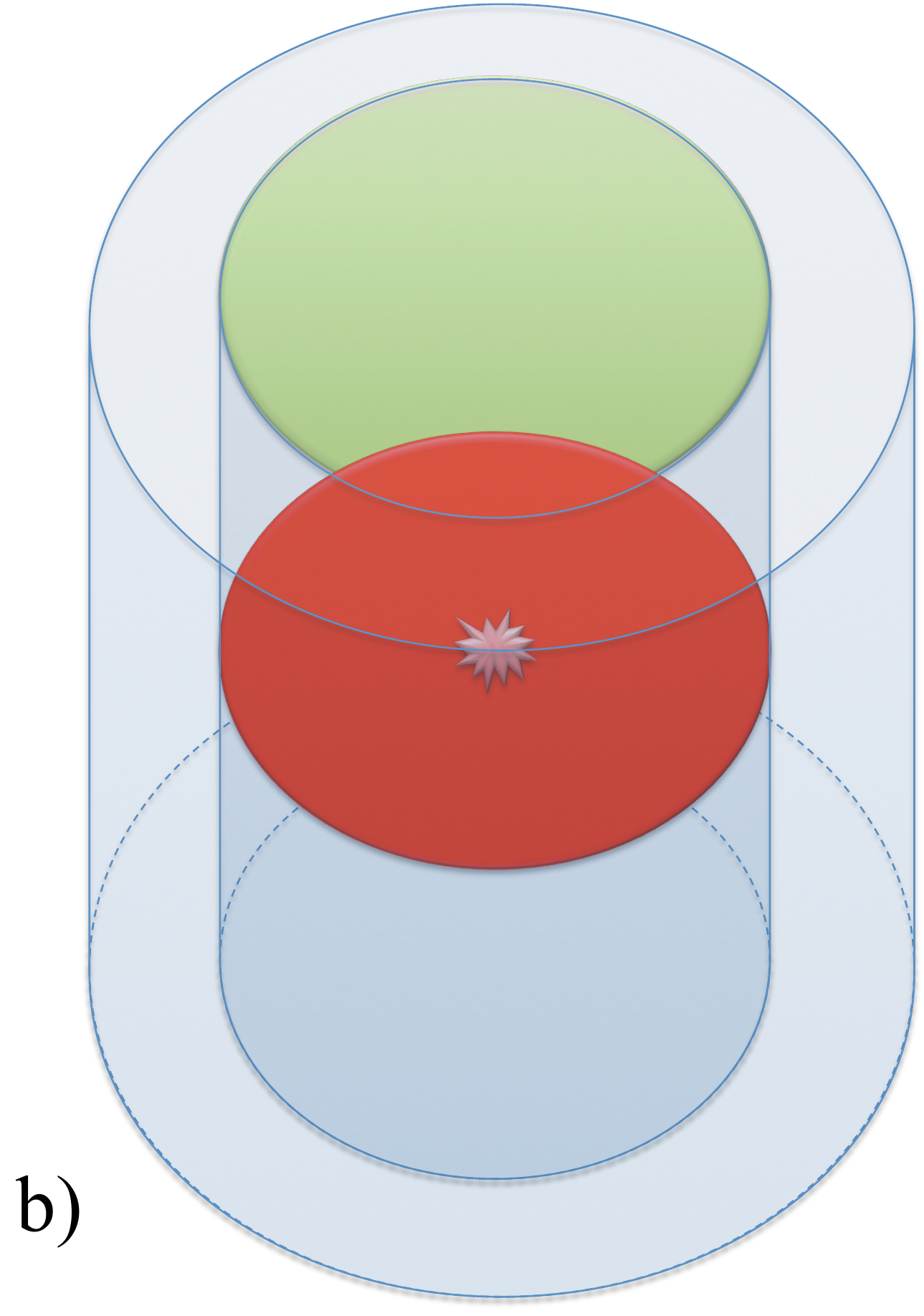} 
\includegraphics[width=3.5cm]{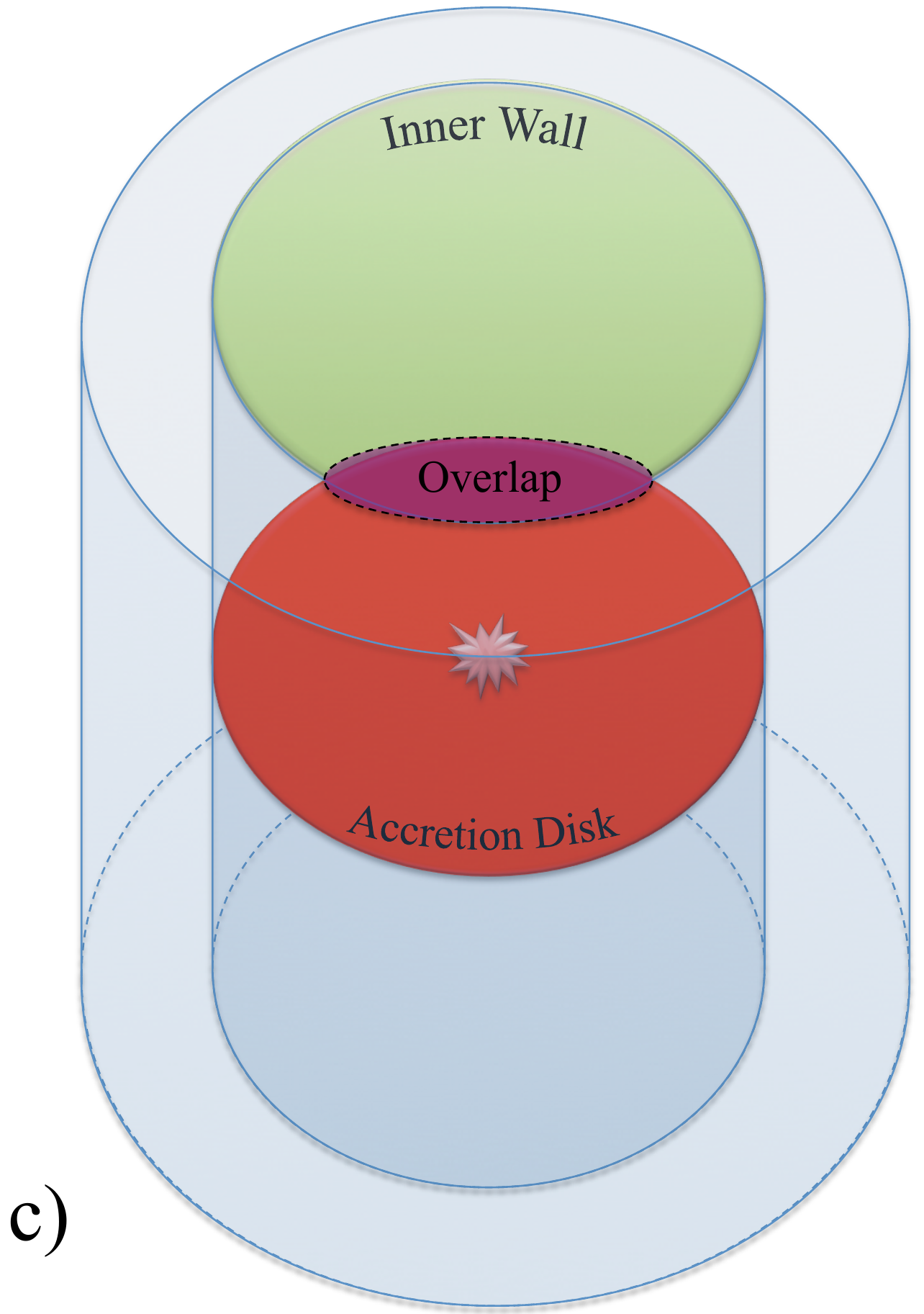} \\
       \caption[Cylinder]{(a) Before the critical angle $\theta_{1}$, the area of the inner wall appears elliptical in shape to the observer as shown in green. (b) Between the critical angles $\theta_{1}$ to $\theta_{2}$, the area of the inner wall, shown in green, and the accretion disk, shown in red, are both seen. (c) The desired area of the inner wall is shown in green, while the overlap is shown in violet. The tilted cylinder model makes the inner wall and the accretion disk appear elliptical in shape.  The overlap of the inner wall and the accretion disk, while not truly elliptical can be well approximated by one, as seen in this figure.}
     \label{fig:cylinder23}
  \end{center}
\end{figure*}

From $\theta_{0}$ to $\theta_{1}$, our model is relatively straightforward. As we are only interested in emission that propagates along the LOS, the emitting area can be calculated as the projection of the curved surface of the cylinder onto the plane of the sky. To the observer, the inner wall appears elliptical in shape as in Figure \ref{fig:cylinder23}a. The area of this ellipse is: $$A_{LOS,1}=\pi r^{2} \sin\theta,$$ where $r$ is the radius of the inner wall of the obscuring region and $\theta$ is the tilt of the cylinder as shown in Figure \ref{fig:cylinder2}. The observed reflected emission from the inner wall will scale with $A_{LOS,1}$.

\subsubsection{Optically Thick Disk Seen: $\theta_{1}<\theta<\theta_{2}$}

\begin{figure}
  \begin{center}
    \leavevmode
      \includegraphics[width=9.5cm]{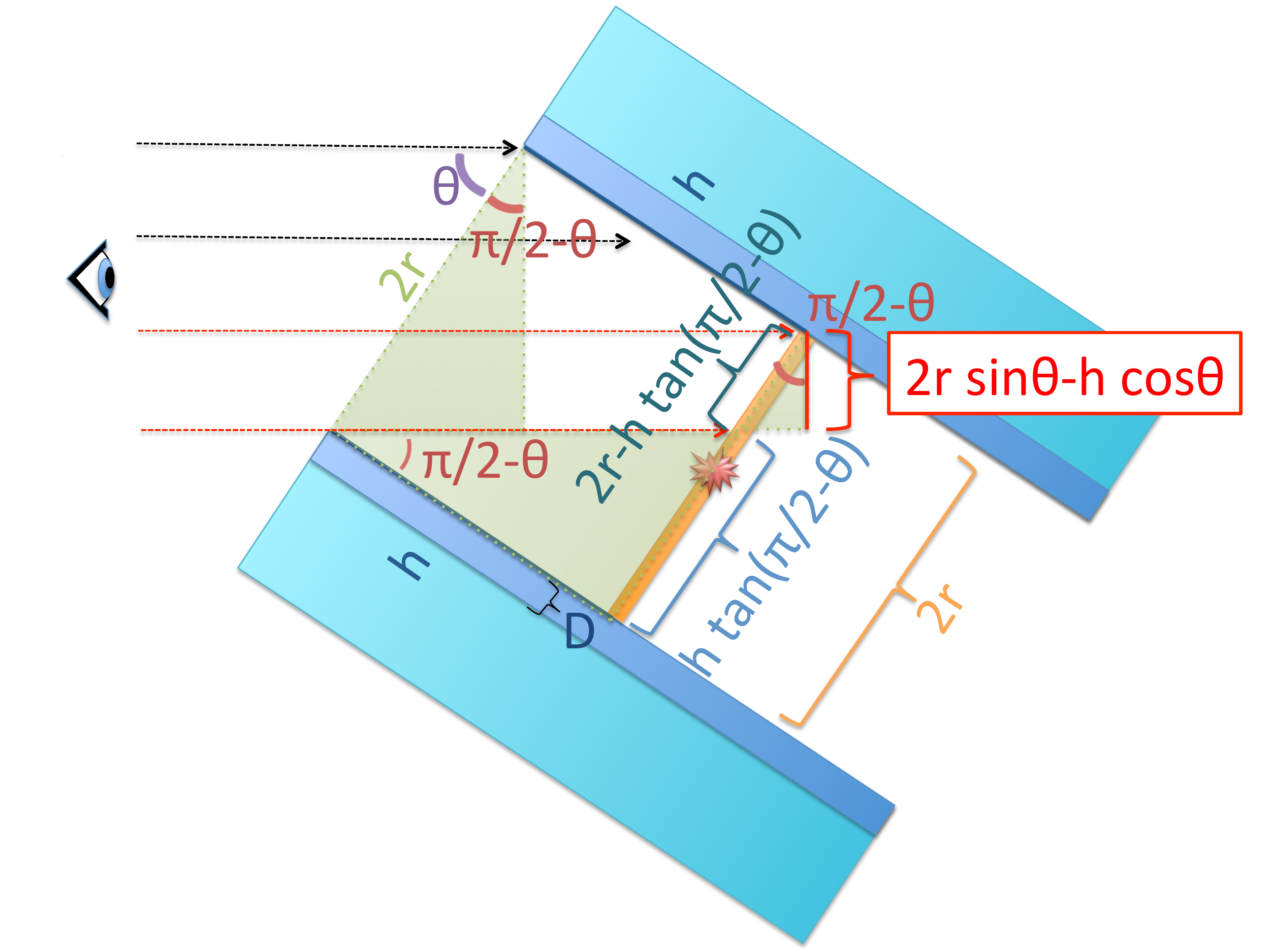} 
       \caption[Cylinder]{A cross section of the cylinder model that can be used to find the area of ~``overlap'' portion of the obscuring disk. The torus is shown in blue with the thickness of the CLiF emitting region, $D$, shaded darker. $D$ is shown for reference, but is not to scale. The obscuring disk is drawn in orange. We can use three similar triangles, each marked with a angle of $\frac{\pi}{2}-\theta$ to find the length of the red line.  This line is equal to twice the semi-minor axis of the~``overlap''.}
     \label{fig:cylinder8}
  \end{center}
\end{figure}

\begin{figure}
  \begin{center}
    \leavevmode
      \includegraphics[width=10cm]{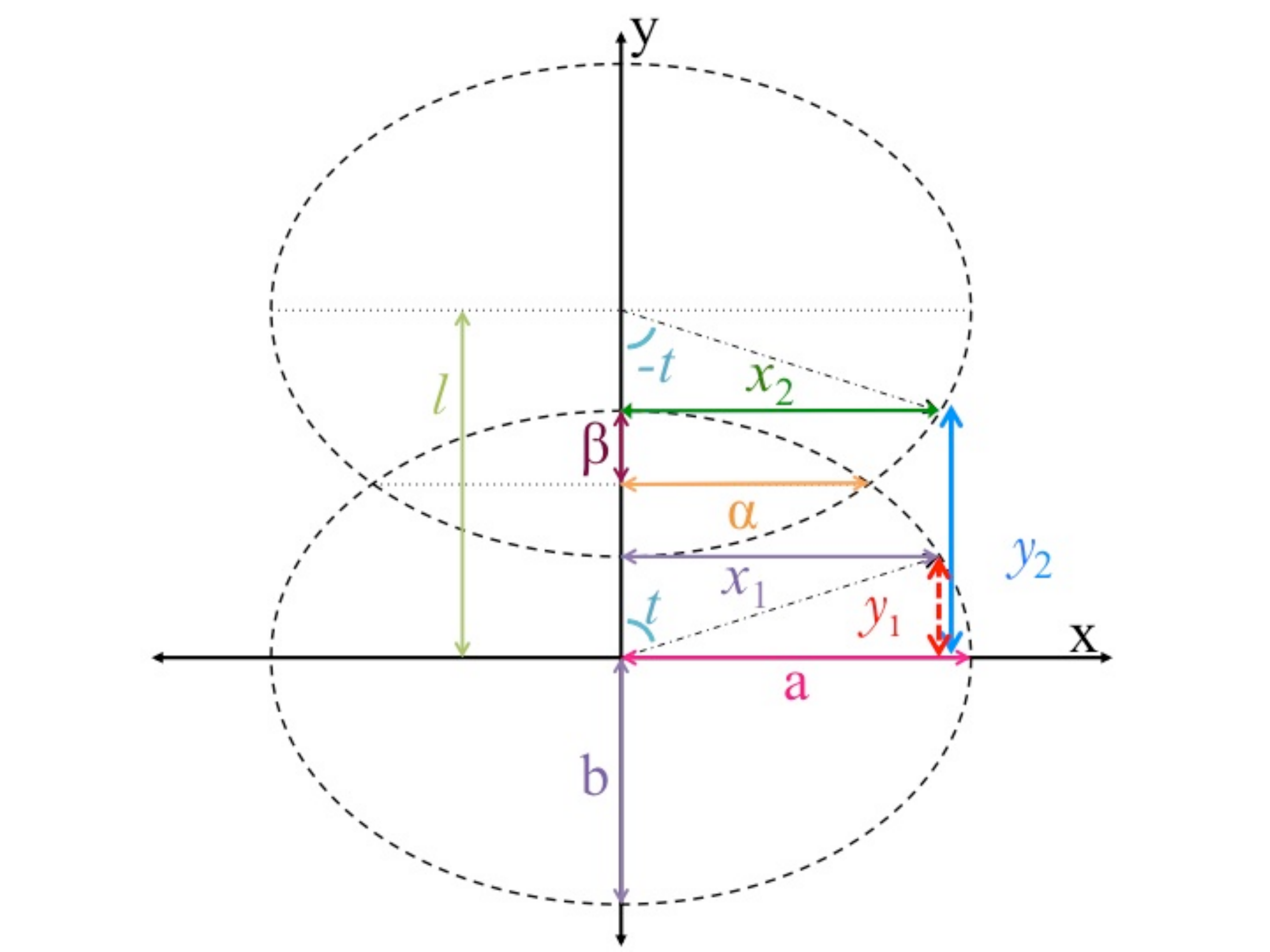} 
       \caption[Cylinder]{The two overlapping ellipses are drawn in Cartesian coordinates, labeled with the relevant variables used to find the area of the ~``overlap'' portion of the obscuring disk. At the special angle of $\theta_{2}$, $y_{1}=y_{2}$.}
     \label{fig:cylinder9}
  \end{center}
\end{figure}

From $\theta_{1}$ to $\theta_{2}$, the model becomes more complex. The observed inner wall of the torus no longer appears as a simple ellipse since part of the accretion disk is revealed, hiding the cylinder wall below the disk plane (Figure \ref{fig:cylinder23}b). To calculate the visible area of the inner wall, we must now subtract the area occulted by the accretion disk. This area is well-approximated (within $\sim$10\% accuracy) by an ellipse (Figure \ref{fig:cylinder23}c.)

To find the area of the overlapping ellipses we need to find the semi-major and semi-minor axes. Using similar triangles, shown in Figure \ref{fig:cylinder8}, we find the semi-minor axis of the overlap, $\beta$, to be 
$$\beta=r\sin\theta-\frac{h}{2}\cos\theta.$$ 
Next, we find the semi-major axis of the overlap, $\alpha$, using the parametric equations for the two ellipses shown in Figure \ref{fig:cylinder9}: 
$$[x_{1},y_{1}]=[2\cos t, \sin t]$$ and $$[x_{2},y_{2}]=[2\cos t, \sin t + l],$$ 
where $l$ is the offset between the centers of the two ellipses determined by the angle of the tilt, $\theta$, and the height, $h$. Setting $y_{1}=y_{2}$, which is true for $\theta_{2}$, and solving for t yields, $$t=\arcsin \frac{l}{2b}.$$ This gives $$x_{1}=x_{2}=a\cos(\arcsin \frac{l}{2b})=a\sqrt{1-\frac{l^{2}}{4b^{2}}}.$$ With $a=r$, $b=r\sin\theta$, $l=h\cos \theta$, we find that the semi-major axis has a length of $$\alpha=r\sqrt{1-\frac{h^{2}}{4r^{2}}\cot^{2}\theta}.$$ Next, we can calculate the area of the overlap to be $$A_{ovlp}=\pi\alpha\beta.$$ Hence the visible area of the inner torus wall is $$A_{LOS,2}=\pi r^{2} \sin\theta-A_{ovlp}$$ from $\theta_{1}$ to $\theta_{2}$. The observed reflected emission from the inner wall will scale with $A_{LOS,2}$.

%%%%%%%%%%%%%%%%%%%%%%%%%%%%%%%%%%%%%%%%%%%%%%%%%%%%%%%%%%%%%%%%%%%%%%%%%%%%%%%%%

\subsubsection{Neglecting Accretion Disk Emission}

So far we have ignored the emission from the accretion disk. There will be an angle at which disk emission begins to become comparable to the [\ion{Fe}{7}] luminosity. We can find this smaller angle, $\theta_{2b}$, by calculating how the temperature of the accretion disk varies with radius. Assuming that the accretion disk is a standard sum of black bodies emitter \citep{Frank}: 

$$T=T_{*}\bigg(\frac{R}{R_{*}}\bigg)^{-3/4}$$

$$R_{*}=\frac{2GM}{c^2}$$

$$T_{*}=\bigg(\frac{3GM\dot{M}}{8\pi R_{*}^3\sigma}\bigg)^{1/4}.$$ 
At our reference r=~1pc, the temperature of the accretion disk at its outer edge is 40 K for M=$10^7$M$_{\odot}$ and $\dot{\text{M}}=1$M$_{\odot}/$yr. A black body at 40 K peaks in the IR at 77$\mu$m, which does not coincide with [\ion{Fe}{7}]$\lambda$6087 emission. 

A temperature of 1000 K peaks at 3$\mu$m and is 100 times fainter at 0.6$\mu$m. Conservatively, the accretion disk emission can no longer be ignored for $T$>1000 K. Using $T$=1000 K at most, the accretion disk is bright enough in the optical band to swamp inner wall emission lines only when $R$$<$0.018 pc as $T(R)\propto R^{-3/4}$. This restriction further constrains the observing bound as $\theta_{2b}\approx \arctan(\frac{h}{1.018r}).$ As this new value differs by less than $2\%$ from $\theta_{2}=\arctan(\frac{h}{r})$, we shall simply use the original definition.

\subsubsection{Area-Angle Dependence}

\begin{figure}
  \begin{center}
    \leavevmode
     \includegraphics[width=9.5cm]{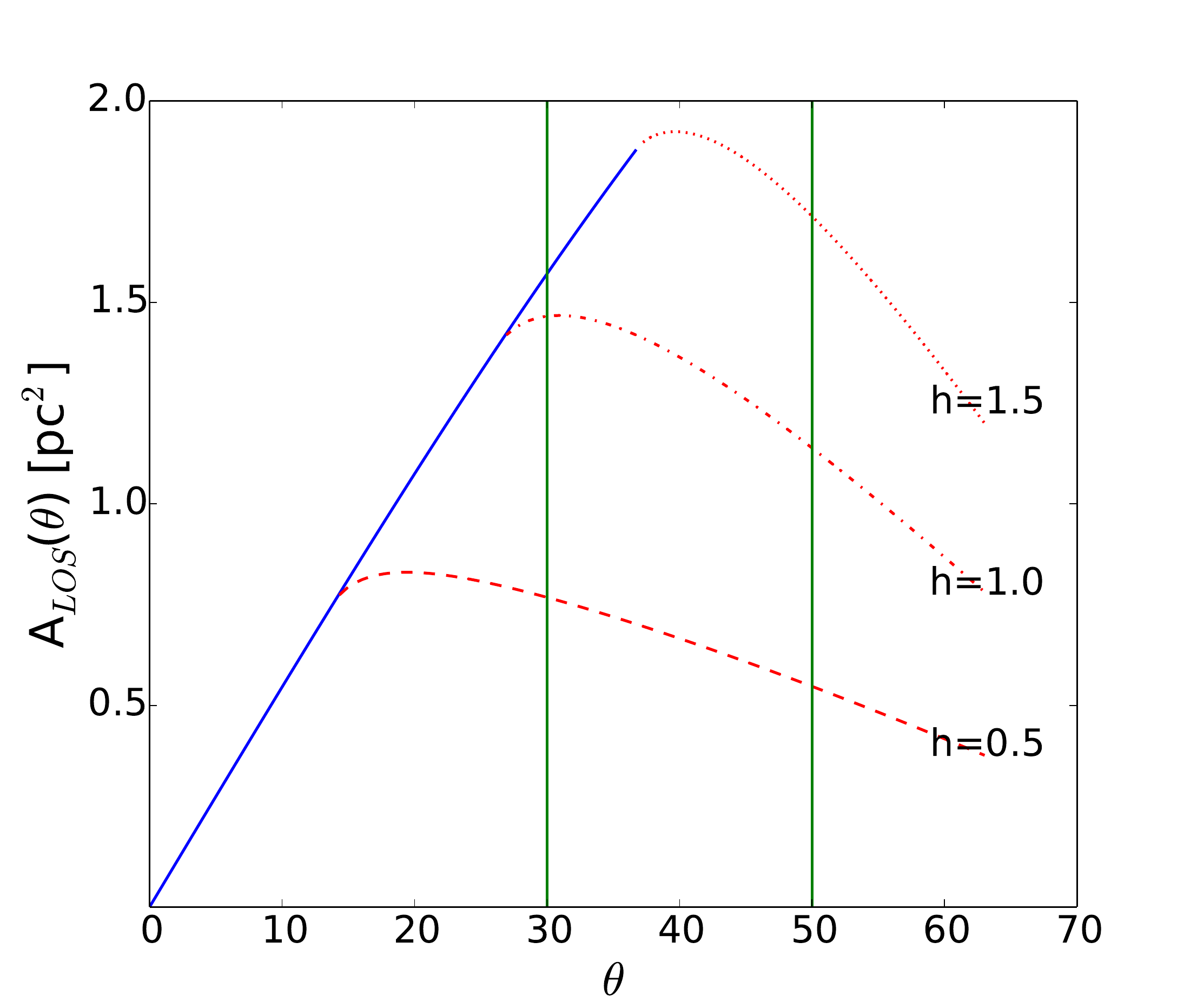}
       \caption[Cylinder]{The dependence of the revealed area, A$_{LOS}$, of the inner wall on the angle, $\theta$. The blue line shows the case in which the accretion disk is hidden from view, while the red dashed and dotted lines shows when both the inner wall and the accretion disk are visible. From bottom to top, the red lines show the cases when r=1.0pc and h=0.5pc (dashed), h=1.0pc (dot-dashed), and h=1.5pc (dotted). According to the WISE-angle relation from \citet{Fischer}, \citet{Rose2015b} found that CLiF type 2 AGN are in the region between $\sim$30$^{\circ}$ and $\sim$50$^{\circ}$ degrees, shown by vertical green lines. To consider the case when the disk does not obscure the lower half of the inner wall, the inner wall area will follow a curve for $A_{LOS}$ that corresponds to the case with double its height, $h$. For instance, compare the examples of $h=0.5$ and $h=1.0$. $h=0.5$ could be with an obscuring accretion disk and $h=1.0$ could be for the same torus, but with no accretion disk.}
     \label{fig:area}
  \end{center}
\end{figure}

The emitting area from the inner wall is described by:
\begin{align}
    \scriptstyle{A_{LOS}(\theta) =} \left\{
     \begin{array}{lr}
       \pi r^{2} \sin\theta, &  \scriptstyle{0<\theta<\theta_{1}}\\
       \scriptstyle{\pi r^{2} \sin\theta-\pi(r\sin\theta-\frac{h}{2}\cos\theta)(r\sqrt{1-\frac{h^{2}}{4r^{2}}\cot^{2}\theta})}, &  \scriptstyle{\theta_{1}<\theta<\theta_{2}}
     \end{array}
   \right.
\end{align}

Figure \ref{fig:area} shows how A$_{LOS}$ changes as a function of $\theta$. Cases for the three adopted values of $h/r$ are shown. The solid blue line is for the area when $0<\theta<\theta_{1}$ and is the same for all values of $h/r$ until their corresponding value of $\theta_{1}$. The red lines show where the area differs for three cases within $\theta_{1}<\theta<\theta_{2}$. The red lines are for r=1.0pc and h=0.5pc (dashed), h=1.0pc (dot-dashed), and h=1.5pc (dotted). Figure \ref{fig:area} can also be used to see how the area of the inner wall changes if there is no equatorial opaque disk so that the lower half of the inner wall is not obscured. In this case, the area follows the blue curve and $A_{LOS}=\pi r^{2} \sin\theta$ until the lower edge of the torus is reached. At this point, we are essentially ``seeing through'' the torus, and the observed inner wall area now decreases. We can again treat this as an ellipse blocking out an ellipse as we did with the obscuring disk. The curve appears the same as that for a torus of double its height with an obscuring disk. For instance, in Figure \ref{fig:area}, we can consider the examples of $h=0.5$ and $h=1.0$. The example with $h=0.5$ could represent a torus with an obscuring disk, while the curve for $h=1.0$ is equivalent to the same torus of $h=0.5$, but with no obscuring disk.

The vertical green lines correspond to 30$^{\circ}$ and 50$^{\circ}$, the range of CLiF type 2 AGN found by \citet{Rose2015b} using the WISE-angle relation from \citet{Fischer}. The best agreement of these two criteria is for $h/r$ = 1.5.

%%%%%%%%%%%%%%%%%%%%%%%%%%       Strength of [FeVII]$\lambda$6087      %%%%%%%%%%%%%%%%%%%%%%%

\section{Emissivity of FHILs} \label{FHILs}

To relate the tilt of the torus to the strength of the FHILs, we need to calculate their emissivity as a function of $\theta$. We take the [\ion{Fe}{7}]$\lambda$6087 emission line as an example as it is typically the most prominent of the FHILs lines in AGN. It also has one of the longest optical [\ion{Fe}{7}] wavelengths, making it relatively unaffected by dust extinction. The same treatment may be applied to other emission lines from the CLiF region such as [\ion{Fe}{6}] and [\ion{Ne}{5}]. Using {\tt Cloudy} photoionization models, \citet{Rose11} showed that they arise at similar ionization parameters and densities. The [\ion{Fe}{7}] luminosity is \citep{Nussbaumer}:

$$F(\lambda_{ij})=\int_{V}\varepsilon_{ij}~dV~~[\text{erg/s}]$$where $F(\lambda_{ij})$ is the detected luminosity from level $j=3f$ to level $i=1d$ in units of ergs s$^{-1}$. V is the volume of the emission region and $\varepsilon_{ij}$ is the emissivity of the transition:

$$\varepsilon_{ij}=N_{j}(\text{Fe}^{6^{+}})A_{ji}h\nu_{ij},$$ where $N_{j}(\text{Fe}^{6^{+}})$ is the number density of Fe$^{6^{+}}$ ions in level j, $A_{ji}$ is the transition probability, and $h\nu_{ij}$ is the photon energy of the transition  \citep{Nussbaumer}.

$N_{j}(\text{Fe}^{6^{+}})$ can be calculated from the number of Fe$^{6^{+}}$ ions, the probability that the electron is in level j, and the number density of iron atoms.  The number density of Fe$^{6^{+}}$ ions can be calculated from the hydrogen density of the CLiF emitting region and the cosmic abundance of iron.  Using {\tt Cloudy}  c13.03 \citep{Ferland}, \citetalias{Rose} found N$_{H}$=$10^{5.5}$ to $10^{7}$ cm$^{-3}$. For $\text{Fe/H}=4\e{-5}$ \citep{Anders} and this range of values of N$_{H}$, we can calculate the number of ions per cm$^{3}$ (Table \ref{table:MultiTable}.).

%%%%%%%%%%%%%%%%%%%%%%%%%%%%%%%%%%%%%%%%%%%%%%%%%%%%%%%%%%%%%%%%%%%%%%%%%%%%%%%%%

\begin{table}
\centering
\caption{}
\begin{tabular}{l|cccc}
\toprule[0.75pt]
\toprule[0.25pt]

 \multicolumn{1}{c|}{\head{}} & \multicolumn{4}{c}{\head{log(U)}}\\
\multicolumn{1}{c|}{\head{}} &\multicolumn{2}{c}{\head{-0.5}}&\multicolumn{2}{c}{\head{-1.0}}\\[2pt]\cline{2-5}\\

\multicolumn{1}{c|}{{\textbf{}}} & \multicolumn{4}{c}{\head{log(n$_{\text{H}}$)}}\\[3pt]
 \multicolumn{1}{c|}{\head{}} & \multicolumn{1}{c}{\head{6.5}}& \multicolumn{1}{c}{\head{7.5} }& \multicolumn{1}{c}{\head{6.5}}& \multicolumn{1}{c}{\head{7.5}}\\
\hline
Fe$^{6^{+}}$[cm$^{-3}]$   & $127$	& $400$& $127$	& $400$ \\
Fe$^{6^{+}}$ fraction   & $$0.0136$$	& $$0.0136$$& $$0.0234$$& $0.0234$ \\
$N_{j}(\text{Fe}^{6^{+}})$ [cm$^{-3}$]   & $0.34$	& $1.1$&  $0.59$		& $1.9$\\
$\varepsilon_{ij}~[10^{-13}~\text{erg}~\text{s}^{-1}~\text{cm}^{-3}]$    & $6.5$	& $20$&  $11$	& $35$ \\
D [$10^{15}$~cm]   & $43$		& $14$&  $8.9$		& $2.0$ \\
D/r    & $0.0139$		& $0.0045$& $0.0029$		& $0.0006$ \\
\hline
\end{tabular}
\label{table:MultiTable}
\end{table}

%%%%%%%%%%%%%%%%%%%%%%%%%%%%%%%%%%%%%%%%%%%%%%%%%%%%%%%%%%%%%%%%%%%%%%%%%%%%%%%%%

The probability that the electron is in level j is calculated to be 0.19, assuming that the electron has fallen into the j level from an upper level, rather than being excited directly to that level \citep{Nussbaumer}. The number density ratio of iron for different values of the ionization parameter, U, can be calculated using {\tt Cloudy}  (Table \ref{table:MultiTable}.) This allows us to calculate $N_{j}(\text{Fe}^{6^{+}})$ (Table \ref{table:MultiTable}). This is an upper limit, assuming that all transitions are from j to i.

Independent of the values of the other parameters, $A_{ji}=0.577$ \citep{Nussbaumer} and $h\nu_{ij}=3.26\e{-12}$ erg~s. Using these values, we can numerically evaluate the emissivity (see Table \ref{table:MultiTable}.) 

The volume, V, of the observed FHILs region depends on $\theta$, N$_{H}$, and U.  The values of the depth, $D$, depend on both U and N$_{H}$. $D$ is shown in Figures \ref{fig:cylinder1}, \ref{fig:cylinder2}, and \ref{fig:cylinder8} as the darker shaded part of the obscuring torus' inner wall. However, it is not shown to scale. The value of $D$ for various combinations of U and N$_{H}$ are given in Table \ref{table:MultiTable}. Table \ref{table:MultiTable} also shows that there are trends with both density and the ionization parameters. The higher ionization gives larger physical depth, as expected in all the relevant cases $D/r$ is small, $<1.5\%$.

%%%%%%%%%%%%%%%%%%%%%%%%%%%%%%%%%%%%%%%%%%%%%%%%%%%%%%%%%%%%%%%%%%%%%%%%%%%%%%%%%

\begin{table}
\centering
\caption{$\text{Predicted L([\ion{Fe}{7}]}\lambda6087)~[10^{40}~\text{erg}~\text{s}^{-1}~\text{cm}^{-3}]$}
\begin{tabular}{cc|cccccc}
\toprule[0.75pt]
\toprule[0.25pt]
 \multicolumn{1}{c}{\head{}} && \multicolumn{6}{c}{\head{log(n$_{\text{H}}$)}}\\
\multicolumn{1}{c}{} && \multicolumn{3}{c}{\head{6.5}} & \multicolumn{3}{c}{\head{7.0}}\\ \cline{3-8}

\multicolumn{1}{c}{} && \multicolumn{6}{c}{\head{h/r, r=1\text{pc}}}\\
\multicolumn{1}{c}{{\textbf{log(U)}}} &\multicolumn{1}{c|}{{\textbf{$\theta$}}}& \multicolumn{1}{c}{\head{0.5}}& \multicolumn{1}{c}{\head{1.0}} & \multicolumn{1}{c}{\head{1.5}}&\multicolumn{1}{c}{\head{0.5}}& \multicolumn{1}{c}{\head{1.0}} & \multicolumn{1}{c}{\head{1.5}}\\
\hline
%\toprule[0.25pt]
        	&	\multicolumn{1}{|c|}{30$^{\circ}$}   & 8.6 & 17 & 26 & 8.8 & 18 & 20   \\[2pt]
  -0.5      	&	\multicolumn{1}{|c|}{40$^{\circ}$}   & 7.6 & 15 & 23 & 7.8  & 16 & 23   \\[2pt]
        	&	\multicolumn{1}{|c|}{50$^{\circ}$}   & 6.4 & 13 & 19 & 6.5 & 13 & 20  \\[2pt]
\hline 
        	&	\multicolumn{1}{|c|}{30$^{\circ}$}    & 3.0 & 6.1 & 7.0 & 2.2 &  4.3 & 5.0   \\[2pt]
  -1.0      	&	\multicolumn{1}{|c|}{40$^{\circ}$}    & 2.7 & 5.4 & 8.1 & 1.9 & 3.8 & 5.7   \\[2pt]
        	&	\multicolumn{1}{|c|}{50$^{\circ}$}   & 2.3  & 4.5 & 6.8 & 1.6 & 3.2 & 4.8  \\[2pt]
\hline

\end{tabular}
\label{table:FLambdaTableobs}
\end{table}

%%%%%%%%%%%%%%%%%%%%%%%%%%%%%%%%%%%%%%%%%%%%%%%%%%%%%%%%%%%%%%%%%%%%%%%%%%%%%%%%%
\begin{table*}
\tiny

\centering
 \begin{threeparttable}

\caption{$\text{Observed L([\ion{Fe}{7}]}\lambda6087)~[10^{40}~\text{erg}~\text{s}^{-1}]$}
\begin{tabular}{ccccccccccc}
\toprule[0.75pt]
\toprule[0.25pt]
\multicolumn{1}{c}{\head{Source}} & \multicolumn{1}{c}{\head{L $[10^{40}~\text{erg}~\text{s}^{-1}]$}$\tnote{a}$}& \multicolumn{1}{c}{\head{log(U)$_{\text{min}}$$\tnote{a}$}}& \multicolumn{1}{c}{\head{log(U)$_{\text{max}}$$\tnote{a}$}}& \multicolumn{1}{c}{\head{log(n$_{\text{H}}$)$_{\text{max}}$$\tnote{a}$}}& \multicolumn{1}{c}{\head{log(n$_{\text{H}}$)$_{\text{max}}$$\tnote{a}$}}&\multicolumn{1}{c}{\head{[W2-W4]$\tnote{b}$}}&\multicolumn{1}{c}{\head{$\theta$$\tnote{c}$}}&\multicolumn{1}{c}{\head{$h/r$$\tnote{d}$}}&\multicolumn{1}{c}{\head{R$_{sub}$$\tnote{a}$}}&\multicolumn{1}{c}{\head{h$_{implied}$}}\\
\hline\\[-5pt]
  Mrk 1388  	&	$0.10\pm0.01$   	& -2.5 	& -1.5 	& 3.5	&4.5 &$5.80\pm0.04$ &$14.0\pm2.7$&0.3 &$0.53\pm0.03$ & 0.16\\
  III Zw 77      	&	$2.6\pm0.3$   	& -1.5	&-1.0 	&3.0		&4.0 &\nodata&\nodata&\nodata &$1.09\pm0.06$ & \nodata\\
  J1241+44	&	$0.23\pm0.02$ 	& -2.5 	& -1.0  	&4.0		&4.5 & $6.04\pm0.13$&$30.0\pm8.7$&0.6 &$0.16\pm0.02$ & 0.10\\
  J1641+43      &	$15\pm4$  		&-2.5 	& 0  		&3.5		&4.5 &$6.07\pm0.06$&$32.0\pm4.0$&0.6 &$2.43\pm0.14$ & 1.46\\
 Q1131+16      &	$21\pm1$ 		&-2.0 	& 0  		&3.5		&4.5  &$5.34\pm0.07$&$-16.7\pm4.7$&\nodata &$1.51\pm0.09$ &\nodata\\
 Tololo 0109-383  &$0.22\pm0.03$	&  -1.5 	& -1.0  	&3.0		&4.0	&$4.95\pm0.03$&$-42.7\pm2.0$&\nodata &$0.53\pm0.02$ &\nodata\\
\\[-6pt]
\hline\\[-2pt]
\end{tabular}
\label{table:CLiFData}
        \begin{tablenotes}
            \item[a] \citetalias{Rose}. \item[b] \cite{Rose2015b}. \item[c] Using the best-fit line [W2-W4]=0.015$\theta$+5.59 from \cite{Rose2015b}. \item[d] $h/r$ is calculated under the (large) assumption that $\theta=\theta_{2} $.
        \end{tablenotes}
\end{threeparttable}
\end{table*}

%%%%%%%%%%%%%%%%%%%%%%%%%%%%%%%%%%%%%%%%%%%%%%%%%%%%%%%%%%%%%%%%%%%%%%%%%%%%%%%%%

F$(\lambda_{ij})$ gives the luminosity emitted from the entire surface area of the FHILs emitting region, A$_{tot}$. We need the fraction of the luminosity along the LOS. This fraction is simply the area calculated in \S2.1, now called A$_{LOS}$, divided by A$_{tot}$:

$$F(\lambda_{ij})_{pred}=\frac{F(\lambda_{ij})~\text{A}_{LOS}}{\text{A}_{tot}}.$$

The small physical depth of the FHIL region ($D$) allows us to ignore the surface area from the thickness of the emitting region. For instance, even with the largest physical value for $D$ of $4.3\e{16}$ cm, it only contributes about $\sim$ 1.0\% of $\text{A}_{tot}$. Therefore, we simply use the surface area of the back and front of the emitting volume. 

For $\theta_{0}<\theta<\theta_{1}$:

$$V=\frac{1}{2}D\pi^{2}r^2{}\tan\theta$$ 

$$\text{A}_{tot}=\pi^{2}r^{2}\tan\theta$$

$$F(\lambda_{ij})_{pred}=\frac{1}{2}\varepsilon_{ij}D\pi r^{2}\sin\theta.$$ Thus, for any object with given $D$ and $r$, the observed luminosity, $F(\lambda_{ij})_{pred}$, varies only with $\sin\theta$.

Then, the predicted luminosity can be described by:

\begin{align}
   \scriptstyle{F(\lambda_{ij})_{pred}=} \left\{
     \begin{array}{lr}
        \scriptstyle{\frac{1}{2}\varepsilon_{ij}D\pi r^{2}\sin\theta,~ \theta_{0}<\theta<\theta_{1}}\\
       \scriptstyle{\scriptstyle{\frac{1}{2}\varepsilon_{ij}D\pi r\big[r\sin\theta-(r\sin\theta-\frac{h}{2}\cos\theta)\sqrt{1-\frac{h^{2}}{4r^{2}}\cot^{2}\theta}\big]}},~\theta_{1}<\theta<\theta_{2}
     \end{array}
   \right.
\end{align} 

Selected values for the predicted luminosity can be found in Table \ref{table:FLambdaTableobs} for these inclinations.  Here we keep to our values of $h/r$ = 0.5, 1.0, 1.5, and $r$ = 1pc, as before. The predicted luminosity is most sensitive to $h/r$ and log(U) (up to factors $\sim$3 and $\sim$4, respectively) and weakly dependent on $\theta$ (up to factors $\sim$3) and log(n$_{\text{H}}$)  (up to factors $\sim$1.5) (Table \ref{table:FLambdaTableobs}).

%%%%%%%%%%%%%%%%%%%%%%%%%%%%%%%%%%%%%%%%%%%%%%%%%%%%%%%%%%%%%%%%%%%%%%%%%%%%%%%%%

\subsection{Comparison with Data}

We can now compare the predicted luminosities with the observed values for the type 2 CLiF AGN from \citetalias{Rose} and \cite{Rose2015b}. Table \ref{table:CLiFData} shows a set of observed values. The predicted values of (1.6-26)\e{40} [erg s$^{-1}$] are in close agreement with the observed values of (0.1-21)\e{40} [erg s$^{-1}$]. 50\% of the objects in Table \ref{table:CLiFData} have [\ion{Fe}{7}] luminosities that are comparable to those in Table \ref{table:FLambdaTableobs}. The rest of the objects in Table \ref{table:CLiFData} have fainter [\ion{Fe}{7}] luminosities by a factor of $\sim$ 10 when compared to the model's, which could be accounted for if the torus were clumpy, as that would lower the surface area.

%%%%%%%%%%%%%%%%%%%%%   DISCUSSION and CONCLUSION  %%%%%%%%%%%%%%%%%%%%%%%%%%%%
\section{Discussion and Conclusions}

We have described a simple geometric model for [\ion{Fe}{7}]$\lambda$6087 emission from the inner wall of the torus and have predicted L([\ion{Fe}{7}]) for appropriate physical conditions to CLiF AGNs.

\begin{enumerate}

\item We showed the relation between the type 2 : type 1 AGN fraction and the $h/r$ ratio.  Type 2 : type 1 AGN ratios of 1 -- 4 require $h/r$=0.6 -- 1.3, and torus opening angles $\theta_{2} \approx 30^{\circ} - 50^{\circ}$. The type 2 : type 1 AGN fraction is then sensitive to small physical changes, which may explain why observed values differ for differently selected samples.

\item For $h/r$ of $0.5-1.5$, the peak FHIL emission at $\sim$15$^{\circ} - 40^{\circ}$ is comparable to the $\sim$30$^{\circ} - 50^{\circ}$ CLiF region found by \citet{Rose2015b}. This matches better for $h/r$ = 1.5.

  \item This model for CLiF AGN predicts L([\ion{Fe}{7}]) values that match the observed values in 50\% of cases: the remainder being a factor $\sim$10 weaker. We discuss reasons for this discrepancy below.

  \item The model could easily be extended to other FHILs, using the relation between viewing angle and the revealed area of the inner wall. 

\end{enumerate}

However, our proposed structure for the torus has oversimplified several aspects of AGN geometry, which may produce the lower luminosities and inclination angles between the model and observations. There are several examples of these oversimplifications:
\begin{enumerate}
  \item In reality, the shape of the inner wall is unlikely to be strictly perpendicular to the accretion disk. 
  \item The torus is likely inhomogeneous, i.e. clumpy \citep{Nenkova2002,Nenkova2008a,Nenkova2008b}. The clumps would lower the observed surface area, while their distribution could lower the amount of sublimated dust as they are in different regimes for U and T. There could also be density variations among the clouds. Some clouds could have densities lower than n$_{crit} \approx 10^{7.6} $cm$^{-3}$ \citep{Nussbaumer}. 
  \item Other material along our LOS may produce dust extinction. 
\item Variability in the accretion disk luminosity would change the predicted luminosities. 
 \end{enumerate}

These effects all tend to lower the observed [\ion{Fe}{7}] luminosity.

Despite these approximations this simple model does predict to reasonable accuracy both the inclination angles and the FHIL luminosities of the type 2 CLiF AGNs.

The cylinder model can also be applied to explain the anomalous Balmer ratio reported in \citetalias{Rose}. \citetalias{Rose} inferred this anomaly from the implied densities and luminosity distance ranges of both the Balmer H$\alpha$/H$\beta$ and [\ion{Fe}{7}] flux ratios, which are shown to be comparable in \citet{Rose11} and \citetalias{Rose}. This supports the idea that there is a significant contribution to these emission line fluxes from the inner torus wall. Therefore, increasing the observed surface area of the dense inner torus wall will increase the NLR and torus wall H$\alpha$ flux relative to H$\beta$. The cylinder model can be used in a similar treatment as shown for [\ion{Fe}{7}] with H$\alpha$/H$\beta$ to reproduce this result.

As the FHILs have ionization potentials in the soft X-ray band (e.g., [\ion{Fe}{7}] at 99.1 eV), it would be interesting to see if they have unusually strong levels of X-ray flux relative to their UV/optical emission. Predictions of other FHIL luminosities would impose tighter constraints. A larger sample of type 2 CLiF AGNs with measured CLiF luminosities would help test the model. Calculating the expected equivalent widths of [\ion{Fe}{7}] lines may also prove insightful. Imaging the kinematics of the bi-cone regions of type 2 CLiF AGNs, if they have them, would be valuable to test this model by measuring their inclination angles more directly.

\acknowledgments
We thank the referee for their insightful comments, which greatly enhanced this work.

\bibliography{msa}

\begin{thebibliography}{}
\expandafter\ifx\csname natexlab\endcsname\relax\def\natexlab#1{#1}\fi

\bibitem[{{Ahn} {et~al.}(2014){Ahn}, {Alexandroff}, {Allende Prieto}, {Anders},
  {Anderson}, {Anderton}, {Andrews}, {Aubourg}, {Bailey}, {Bastien}, \&
  et~al.}]{Ahn}
{Ahn}, C.~P., {Alexandroff}, R., {Allende Prieto}, C., {et~al.} 2014, \apjs,
  211, 17

\bibitem[{{Anders} \& {Grevesse}(1989)}]{Anders}
{Anders}, E., \& {Grevesse}, N. 1989, \gca, 53, 197

\bibitem[{{Antonucci}(1993)}]{Antonucci}
{Antonucci}, R. 1993, \araa, 31, 473

\bibitem[{{Antonucci} \& {Miller}(1985)}]{AntonucciMiller}
{Antonucci}, R.~R.~J., \& {Miller}, J.~S. 1985, \apj, 297, 621

\bibitem[{{Czerny} \& {Hryniewicz}(2011)}]{Czerny}
{Czerny}, B., \& {Hryniewicz}, K. 2011, \aap, 525, L8

\bibitem[{{Elitzur}(2008)}]{Elitzur}
{Elitzur}, M. 2008, NewAR, 52, 274

\bibitem[{{Fausnaugh} {et~al.}(2015){Fausnaugh}, {Denney}, {Barth}, {Bentz},
  {Bottorff}, {Carini}, {Croxall}, {De Rosa}, {Goad}, {Horne}, {Joner},
  {Kaspi}, {Kim}, {Klimanov}, {Kochanek}, {Leonard}, {Netzer}, {Peterson},
  {Schnulle}, {Sergeev}, {Vestergaard}, {Zheng}, {Zu}, {Anderson}, {Arevalo},
  {Bazhaw}, {Borman}, {Boroson}, {Brandt}, {Breeveld}, {Brewer}, {Cackett},
  {Crenshaw}, {Dalla Bonta}, {De Lorenzo-Caceres}, {Dietrich}, {Edelson},
  {Efimova}, {Ely}, {Evans}, {Filippenko}, {Flatland}, {Gehrels}, {Geier},
  {Gelbord}, {Gonzalez}, {Gorjian}, {Grier}, {Grupe}, {Hall}, {Hicks},
  {Horenstein}, {Hutchison}, {Im}, {Jensen}, {Jones}, {Kaastra}, {Kelly},
  {Kennea}, {Kim}, {Korista}, {Kriss}, {Lee}, {Lira}, {MacInnis},
  {Manne-Nicholas}, {Mathur}, {McHardy}, {Montouri}, {Musso}, {Nazarov},
  {Norris}, {Nousek}, {Okhmat}, {Pancoast}, {Papadakis}, {Parks}, {Pei},
  {Pogge}, {Pott}, {Rafter}, {Rix}, {Saylor}, {Schimoia}, {Siegel}, {Spencer},
  {Starkey}, {Sung}, {Teems}, {Treu}, {Turner}, {Uttley}, {Villforth}, {Weiss},
  {Woo}, {Yan}, \& {Young}}]{Fausnaugh}
{Fausnaugh}, M.~M., {Denney}, K.~D., {Barth}, A.~J., {et~al.} 2015, ArXiv
  e-prints, arXiv:1510.05648

\bibitem[{{Ferland} {et~al.}(1998){Ferland}, {Korista}, {Verner}, {Ferguson},
  {Kingdon}, \& {Verner}}]{Ferland}
{Ferland}, G.~J., {Korista}, K.~T., {Verner}, D.~A., {et~al.} 1998, \pasp, 110,
  761

\bibitem[{{Fischer} {et~al.}(2013){Fischer}, {Crenshaw}, {Kraemer}, \&
  {Schmitt}}]{Fischer13}
{Fischer}, T.~C., {Crenshaw}, D.~M., {Kraemer}, S.~B., \& {Schmitt}, H.~R.
  2013, \apjs, 209, 1

\bibitem[{{Fischer} {et~al.}(2014){Fischer}, {Crenshaw}, {Kraemer}, {Schmitt},
  \& {Turner}}]{Fischer}
{Fischer}, T.~C., {Crenshaw}, D.~M., {Kraemer}, S.~B., {Schmitt}, H.~R., \&
  {Turner}, T.~J. 2014, \apj, 785, 25

\bibitem[{{Frank} {et~al.}(2002){Frank}, {King}, \&
  {Raine}}]{Frank}
{Frank}, J., {King}, A., \& {Raine}, D.~J. 2002, {Accretion Power in
  Astrophysics: Third Edition}, 398

\bibitem[{{Gilli} {et~al.}(2007){Gilli}, {Comastri}, \&
  {Hasinger}}]{Gilli}
{Gilli}, R., {Comastri}, A., \& {Hasinger}, G. 2007, \aap, 463, 79

\bibitem[{{Hao} {et~al.}(2005){Hao}, {Strauss}, {Fan}, {Tremonti}, {Schlegel},
  {Heckman}, {Kauffmann}, {Blanton}, {Gunn}, {Hall}, {Ivezi{\'c}}, {Knapp},
  {Krolik}, {Lupton}, {Richards}, {Schneider}, {Strateva}, {Zakamska},
  {Brinkmann}, \& {Szokoly}}]{Hao}
{Hao}, L., {Strauss}, M.~A., {Fan}, X., {et~al.} 2005, \aj, 129, 1795

\bibitem[{{Komossa} {et~al.}(2008){Komossa}, {Zhou}, \&
  {Lu}}]{Komossa}
{Komossa}, S., {Zhou}, H., \& {Lu}, H. 2008, \apjl, 678, L81

\bibitem[{{Koshida} {et~al.}(2014){Koshida}, {Minezaki}, {Yoshii}, {Kobayashi},
  {Sakata}, {Sugawara}, {Enya}, {Suganuma}, {Tomita}, {Aoki}, \&
  {Peterson}}]{Koshida}
{Koshida}, S., {Minezaki}, T., {Yoshii}, Y., {et~al.} 2014, \apj, 788, 159

\bibitem[{{Lawrence} \& {Elvis}(1982)}]{Lawrence1}
{Lawrence}, A., \& {Elvis}, M. 1982, \apj, 256, 410

\bibitem[{{Lawrence} \& {Elvis}(2010)}]{Lawrence2}
---. 2010, \apj, 714, 561

\bibitem[{{Lira} {et~al.}(2010){Lira}, {Ar{\'e}valo}, {Uttley}, {McHardy}, \&
  {Breedt}}]{Lira}
{Lira}, P., {Ar{\'e}valo}, P., {Uttley}, P., {McHardy}, I., \& {Breedt}, E.
  2010, in IAU Symposium, Vol. 267, Co-Evolution of Central Black Holes and
  Galaxies, ed. B.~M. {Peterson}, R.~S. {Somerville}, \& T.~{Storchi-Bergmann},
  90--95

\bibitem[{{Malkan}(1983)}]{Malkan1983}
{Malkan}, M.~A. 1983, \apj, 268, 582

\bibitem[{{Malkan} \& {Sargent}(1982)}]{Malkan1982}
{Malkan}, M.~A., \& {Sargent}, W.~L.~W. 1982, \apj, 254, 22

\bibitem[{{Marin}(2014)}]{Marin}
{Marin}, F. 2014, \mnras, 441, 551

\bibitem[{{McLure} \& {Dunlop}(2002)}]{McLure}
{McLure}, R.~J., \& {Dunlop}, J.~S. 2002, \mnras, 331, 795

\bibitem[{{Murayama} \& {Taniguchi}(1998)}]{Murayama}
{Murayama}, T., \& {Taniguchi}, Y. 1998, \apjl, 497, L9

\bibitem[{{Nagao} {et~al.}(2003){Nagao}, {Murayama}, {Shioya}, \&
  {Taniguchi}}]{Nagao2003}
{Nagao}, T., {Murayama}, T., {Shioya}, Y., \& {Taniguchi}, Y. 2003, \aj, 125,
  1729

\bibitem[{{Nagao} {et~al.}(2001){Nagao}, {Murayama}, \&
  {Taniguchi}}]{Nagao2001}
{Nagao}, T., {Murayama}, T., \& {Taniguchi}, Y. 2001, \apj, 549, 155

\bibitem[{{Nagao} {et~al.}(2000){Nagao}, {Taniguchi}, \&
  {Murayama}}]{Nagao2000}
{Nagao}, T., {Taniguchi}, Y., \& {Murayama}, T. 2000, \aj, 119, 2605

\bibitem[{{Nenkova} {et~al.}(2002){Nenkova}, {Ivezi{\'c}}, \&
  {Elitzur}}]{Nenkova2002}
{Nenkova}, M., {Ivezi{\'c}}, {\v Z}., \& {Elitzur}, M. 2002, \apjl, 570, L9

\bibitem[{{Nenkova} {et~al.}(2008{\natexlab{a}}){Nenkova}, {Sirocky},
  {Ivezi{\'c}}, \& {Elitzur}}]{Nenkova2008a}
{Nenkova}, M., {Sirocky}, M.~M., {Ivezi{\'c}}, {\v Z}., \& {Elitzur}, M.
  2008{\natexlab{a}}, \apj, 685, 147

\bibitem[{{Nenkova} {et~al.}(2008{\natexlab{b}}){Nenkova}, {Sirocky},
  {Nikutta}, {Ivezi{\'c}}, \& {Elitzur}}]{Nenkova2008b}
{Nenkova}, M., {Sirocky}, M.~M., {Nikutta}, R., {Ivezi{\'c}}, {\v Z}., \&
  {Elitzur}, M. 2008{\natexlab{b}}, \apj, 685, 160

\bibitem[{{Nussbaumer} {et~al.}(1982){Nussbaumer}, {Storey}, \&
  {Storey}}]{Nussbaumer}
{Nussbaumer}, H., {Storey}, P.~J., \& {Storey}, P.~J. 1982, \aap, 113, 21

\bibitem[{{Pancoast} {et~al.}(2014){Pancoast}, {Brewer}, {Treu}, {Park},
  {Barth}, {Bentz}, \& {Woo}}]{Pancoast}
{Pancoast}, A., {Brewer}, B.~J., {Treu}, T., {et~al.} 2014, \mnras, 445, 3073

\bibitem[{{Pier} \& {Voit}(1995)}]{Pier}
{Pier}, E.~A., \& {Voit}, G.~M. 1995, \apj, 450, 628

\bibitem[{{Risaliti} {et~al.}(1999){Risaliti}, {Maiolino}, \&
  {Salvati}}]{Risaliti}
{Risaliti}, G., {Maiolino}, R., \& {Salvati}, M. 1999, \apj, 522, 157

\bibitem[{{Rose} {et~al.}(2011){Rose}, {Tadhunter}, {Holt}, {Ramos Almeida}, \&
  {Littlefair}}]{Rose11}
{Rose}, M., {Tadhunter}, C.~N., {Holt}, J., {Ramos Almeida}, C., \&
  {Littlefair}, S.~P. 2011, \mnras, 414, 3360

\bibitem[{{Rose} {et~al.}(2015{\natexlab{a}}){Rose}, {Elvis}, \&
  {Tadhunter}}]{Rose}
{Rose}, M., {Elvis}, M., \& {Tadhunter}, C.~N. 2015{\natexlab{a}}, \mnras, 448,
  2900

\bibitem[{{Rose} {et~al.}(2015{\natexlab{b}}){Rose}, {Elvis}, {Crenshaw}, \&
  {Glidden}}]{Rose2015b}
{Rose}, M., {Elvis}, M., {Crenshaw}, M., \& {Glidden}, A. 2015{\natexlab{b}},
  \mnras, 451, L11

\bibitem[{{Shakura} \& {Sunyaev}(1973)}]{Shakura}
{Shakura}, N.~I., \& {Sunyaev}, R.~A. 1973, \aap, 24, 337

\bibitem[{{Shankar} {et~al.}(2013){Shankar}, {Weinberg}, \&
  {Miralda-Escud{\'e}}}]{Shankar}
{Shankar}, F., {Weinberg}, D.~H., \& {Miralda-Escud{\'e}}, J. 2013, \mnras,
  428, 421

\bibitem[{{Shields}(1978)}]{Shields}
{Shields}, G.~A. 1978, \nat, 272, 706

\bibitem[{{Suganuma} {et~al.}(2006){Suganuma}, {Yoshii}, {Kobayashi},
  {Minezaki}, {Enya}, {Tomita}, {Aoki}, {Koshida}, \&
  {Peterson}}]{Suganuma}
{Suganuma}, M., {Yoshii}, Y., {Kobayashi}, Y., {et~al.} 2006, \apj, 639, 46

\bibitem[{{Wang} {et~al.}(2012){Wang}, {Zhou}, {Komossa}, {Wang}, {Yuan}, \&
  {Yang}}]{Wang}
{Wang}, T.-G., {Zhou}, H.-Y., {Komossa}, S., {et~al.} 2012, \apj, 749, 115

\bibitem[{{York} {et~al.}(2000){York}, {Adelman}, {Anderson}, {Anderson},
  {Annis}, {Bahcall}, {Bakken}, {Barkhouser}, {Bastian}, {Berman}, {Boroski},
  {Bracker}, {Briegel}, {Briggs}, {Brinkmann}, {Brunner}, {Burles}, {Carey},
  {Carr}, {Castander}, {Chen}, {Colestock}, {Connolly}, {Crocker}, {Csabai},
  {Czarapata}, {Davis}, {Doi}, {Dombeck}, {Eisenstein}, {Ellman}, {Elms},
  {Evans}, {Fan}, {Federwitz}, {Fiscelli}, {Friedman}, {Frieman}, {Fukugita},
  {Gillespie}, {Gunn}, {Gurbani}, {de Haas}, {Haldeman}, {Harris}, {Hayes},
  {Heckman}, {Hennessy}, {Hindsley}, {Holm}, {Holmgren}, {Huang}, {Hull},
  {Husby}, {Ichikawa}, {Ichikawa}, {Ivezi{\'c}}, {Kent}, {Kim}, {Kinney},
  {Klaene}, {Kleinman}, {Kleinman}, {Knapp}, {Korienek}, {Kron}, {Kunszt},
  {Lamb}, {Lee}, {Leger}, {Limmongkol}, {Lindenmeyer}, {Long}, {Loomis},
  {Loveday}, {Lucinio}, {Lupton}, {MacKinnon}, {Mannery}, {Mantsch}, {Margon},
  {McGehee}, {McKay}, {Meiksin}, {Merelli}, {Monet}, {Munn}, {Narayanan},
  {Nash}, {Neilsen}, {Neswold}, {Newberg}, {Nichol}, {Nicinski}, {Nonino},
  {Okada}, {Okamura}, {Ostriker}, {Owen}, {Pauls}, {Peoples}, {Peterson},
  {Petravick}, {Pier}, {Pope}, {Pordes}, {Prosapio}, {Rechenmacher}, {Quinn},
  {Richards}, {Richmond}, {Rivetta}, {Rockosi}, {Ruthmansdorfer}, {Sandford},
  {Schlegel}, {Schneider}, {Sekiguchi}, {Sergey}, {Shimasaku}, {Siegmund},
  {Smee}, {Smith}, {Snedden}, {Stone}, {Stoughton}, {Strauss}, {Stubbs},
  {SubbaRao}, {Szalay}, {Szapudi}, {Szokoly}, {Thakar}, {Tremonti}, {Tucker},
  {Uomoto}, {Vanden Berk}, {Vogeley}, {Waddell}, {Wang}, {Watanabe},
  {Weinberg}, {Yanny}, {Yasuda}, \& {SDSS Collaboration}}]{York}
{York}, D.~G., {Adelman}, J., {Anderson}, Jr., J.~E., {et~al.} 2000, \aj, 120,
  1579

\end{thebibliography}

\end{document}